\documentclass[floatfix,amsmath,amsfonts,onecolumn]{revtex4}  
\usepackage{enumitem}
\usepackage{amsmath}
\usepackage{graphicx}
\usepackage{epsfig}
\usepackage{bm}
\usepackage{bbold}
\usepackage{amssymb}
\usepackage{color}

\newcommand{\PT}{{\cal PT}}

\newcommand{\cL}{{\cal L}}

\newcommand{\tx}{\tilde{x}}
\newcommand{\tu}{\tilde{u}}
\newcommand{\tv}{\tilde{v}}

\newcommand{\ra}{\rangle}
\newcommand{\la}{\langle}
\newcommand{\uox}{u_{0,x}}

\newcommand{\sech}{{\operatorname{sech}}}

\newcommand{\vep}{\varepsilon}
 
\newcommand{\rev}[1]{\textcolor{black}{#1}}

\begin{document}

\title{On \rev{non-}existence of continuous families of stationary nonlinear modes for a class of  complex potentials}

\author{D. A. Zezyulin$^{1}$\footnote{email: d.zezyulin@gmail.com},  A. O. Slobodyanyuk$^{2}$, G. L. Alfimov$^{2}$}

\affiliation{$^{1}$ ITMO University, St.~Petersburg 197101, Russia\\
	$^{2}$  Moscow Institute of Electronic Engineering, Zelenograd, Moscow, 124498, Russia}

\date{\today}

\begin{abstract}
There are two cases when the nonlinear  Schr\"odinger equation (NLSE)  with an external complex potential   is well-known to support continuous families of  localized stationary    modes:  the  ${\cal PT}$-symmetric potentials and the Wadati potentials. Recently Y.~Kominis and coauthors [Chaos, Solitons and Fractals, {\bf 118}, 222-233  (2019)]  have suggested that the  continuous families  can be also found  in  complex potentials of the form $W(x)=W_{1}(x)+iCW_{1,x}(x)$, where $C$ is an arbitrary real and $W_1(x)$ is a real-valued and bounded differentiable function.  Here we   study    in detail   nonlinear stationary modes  that emerge   in complex potentials of this type (for brevity,  we call them  {\it W-dW potentials}).   First, we assume that the potential is small  and employ asymptotic methods to construct a family of  nonlinear modes.  Our asymptotic procedure stops at the terms of the $\varepsilon^2$ order, where small $\vep$ characterizes   amplitude of the  potential. We therefore conjecture   that no continuous families of authentic nonlinear modes    exist in this case, but ``pseudo-modes'' that satisfy the equation up to $\varepsilon^2$-error can indeed be found in W-dW potentials. Second, we consider the particular case of a W-dW potential well of finite depth and support our hypothesis with qualitative  and numerical arguments. Third, we simulate the nonlinear dynamics of found pseudo-modes and observe that,   if the amplitude of W-dW potential is small,  then the pseudo-modes are robust and   display  persistent   oscillations around a certain  position predicted by the asymptotic expansion. Finally, we study the authentic stationary modes  which do not form a continuous family, but  exist as isolated points. Numerical simulations reveal dynamical instability of these solutions. \\[2mm]


\end{abstract}

\maketitle

\section{Introduction}\label{sec:Intro}

One of fundamental differences in properties of nonlinear conservative and dissipative systems is the structure of their stationary modes. It is typical of conservative systems to support continuous families of localized  nonlinear modes which result from the combined effect of the  linear broadening, inhomogeneity of the conservative medium,  and the nonlinear self-action.  However, for stationary modes  that   appear in a dissipative medium the situation becomes more complicated, because an additional balance between  gain and  loss of  energy   is required to sustain the steady-state propagation  \cite{AA08,Rosanov09,Kartashov,Malomed14}. In view of this new  requirement, the structure of dissipative stationary modes is usually  scarcer than in the conservative case,  and, instead of continuous families, dissipative stationary  modes  exist only as isolated points.

A prominent  example of this dissimilarity is provided by the nonlinear Schr\"odinger equation (NLSE) with an  additional, real- or complex-valued  potential (alias the Gross-Pitaevskii equation). A spatially one-dimensional  version of  this equation reads
\begin{equation}
\label{Eq:opt-eq}
i\Phi_t +  \Phi_{xx} - W(x) \Phi  + \rev{\sigma}   |\Phi|^2\Phi=0,
\end{equation}
\rev{where $W(x)$ is the potential, and $\sigma$ is the real nonlinearity coefficient.}
Equation~(\ref{Eq:opt-eq}) arises in various areas of present-day   physics. In optics, it describes the laser beam propagation  in nonlinear media with  the refractive index  modulated in the transverse direction \cite{KA03}. In the theory of Bose-Einstein condensate (BEC) Eq.~(\ref{Eq:opt-eq})  models the dynamics of a cigar-shaped cloud of ultracold quantum gas trapped by an external field  \cite{PS03}. Solitary-wave  stationary modes for this equation correspond to the substitution $\Phi(x,t)=e^{i\mu t} \phi(x)$, where $\mu$ is a real parameter and $\phi(x)$ is the localized stationary wavefunction. If  the potential $W(x)$ is {\it real-valued}, then the model is conservative. It is well-known that it supports  continuous one-parametric families of nonlinear modes which can be obtained by the continuous change of $\mu$. This situation has been comprehensively documented for various types of the external potential, including periodic, parabolic, double-well one, and for either   sign in front of the nonlinear term --- see for instance \cite{Kunze1999,Kivshar2001,Peli04,Theo2007,AZ07,Zezyulin08,Pelinovsky2011,Yang2010}  where this aspect has received   the special emphasis. In the meantime,  the situation can change drastically   when one switches from real to {\it complex} potentials in  (\ref{Eq:opt-eq}),   i.e., assumes  $W(x)=W_1(x)+iW_2(x)$. 
In the optical context, the imaginary part of the potential describes the transverse distribution of gain and losses along the guiding medium, and in the BEC theory it demarcates      spatial regions where the particles are absorbed  from or pumped in the condensate. 

The model with a complex potential is no longer conservative, and it is expected {\it in general}  that its  nonlinear modes  will only exist  as some ``isolated'' points  that cannot be continued in $\mu$.  However it is known that  there exist at least {\it two} situations when Eq.~(\ref{Eq:opt-eq}) with a complex potential supports continuous families of stationary modes. The first example  corresponds to {\it ${\cal PT}$-symmetric potentials}   \cite{Chr08,KYZ16,Kivshar15}, when   $W_1(x)$ and $W_2(x)$ are even and odd functions, respectively.  Physically, the continuous families in $\PT$-symmetric potentials can be understood as a result of   the synergy between  symmetry of the potential and   that of    solution itself, which facilitates the gain-and-loss balance. Rigorous analyses of bifurcations of continuous $\PT$-symmetric families have been reported on recently  \cite{Dohnal16,Dohnal20}. The second class of complex potentials that enable continuous families of nonlinear modes   corresponds to the so-called {\it Wadati potentials} \cite{Wadati}, where   real and imaginary parts of $W(x)$ are expressed through an auxiliary real-valued function $w(x)$ as follows 
\begin{gather*}
W_1(x)=-w^2(x),\quad W_2(x)=-w_x(x).
\end{gather*}
Function $w(x)$  is required to be differentiable, but is not supposed to bear any special symmetry.  Existence of continuous families   in Wadati potentials can be qualitatively explained   by the fact that the ODE system that describes the shape of   stationary modes has  a conserved quantity   which effectively decreases the order of  associated dynamical system   \cite{Tsoy14,KZ14}. Formal asymptotic expansions for families of nonlinear modes bifurcating from linear eigenstates of Wadati potentials have been recently obtained in \cite{Yang2021}.

\rev{It should be also added that unusual properties of $\PT$-symmetric and Wadati potentials appear not only for nonlinear modes but in the linear case, too. In particular, complex eigenvalues that eventually emerge in the spectra of the corresponding non-Hermitian Schr\"odinger operators [obtained from Eq.~(\ref{Eq:opt-eq}) with $\sigma=0$] always form complex-conjugate pairs \cite{Most,Nixon}. A generic complex potential does not have this property. }

In  recent paper \cite{KCKFB19} by Y.~Kominis \emph{et al.}, it has been suggested that there exists yet another class of complex potentials that support continuous families of stationary modes.   Real and imaginary parts of those potentials are related as  
\begin{gather}
W_{1,x}(x)=CW_2(x),\label{Eq:RelMain}
\end{gather}
where $C$ is an arbitrary real constant, and the subscript $x$ means the derivative. The relation (\ref{Eq:RelMain}) was arrived in \cite{KCKFB19} at  by means of the Melnikov-vector technique. 
\rev{In contrast to $\PT$-symmetric potentials which relies on the parity of real and imaginary parts of the potential, condition (\ref{Eq:RelMain}) only involves a spatially local relation between the real and imaginary parts. This  can potentially offer  additional   flexibility in the experimental realization of complex potentials of this form. Since the  specific shapes of $W_1(x)$ and $W_2(x)$ are not constrained,  relations (\ref{Eq:RelMain}) can be used to create complex potentials of different forms: localized or extended, periodic or quasiperiodic, etc.}  Peculiar nonlinear dynamics in potentials (\ref{Eq:RelMain}) have been   studied in several earlier studies \cite{Kominis_OC15,Kominis_PRA15}.

\rev{Let us  briefly overview the outcomes of \cite{KCKFB19}. Authors consider Eq.~(\ref{Eq:opt-eq}) with the focusing (attractive) nonlinearity, in the situation when the complex potential is characterized by a  small amplitude proportional to $\vep \ll  1$:  
	\begin{gather}
	\label{Eq:Phi}
	i\Phi_t +\Phi_{xx} - \varepsilon(W_1(x)+iW_2(x)) \Phi  + 2|\Phi|^2\Phi=0.
	\end{gather}
	Stationary modes have  the  form $\Phi(x,t)=e^{i\mu t} \phi(x)$, where $\mu$ is a real coefficient whose optical meaning corresponds to the propagation constant. Continuously-differentiable  function $\phi(x)$    satisfies the solitary-wave boundary conditions
	$\lim_{x\to\infty}\phi(x) = \lim_{x\to-\infty} \phi(x) =0$.  
	The shape of $\phi(x)$ is described by   the stationary equation
	\begin{gather}
	\label{Eq:phi}
	\phi_{xx} - \mu\phi   - \varepsilon(W_1(x)+iW_2(x)) \phi  + 2|\phi|^2\phi=0.
	\end{gather}
	Equation~(\ref{Eq:phi}) is invariant with respect to the phase rotation   $\phi(x)\to\phi(x)e^{i\theta}$, $\theta \in \mathbb{R}$. 
	Separating $\phi(x)$ into real and imaginary parts, $\phi(x)=u(x) + iv(x)$, one transforms   Eq.~(\ref{Eq:phi}) into a system 
	\begin{align}
	&u_{xx}- \mu u+2(u^2+v^2)u-\varepsilon(W_1(x) u-W_2(x)v)=0,\label{Eq:u-intro}\\[2mm]
	&v_{xx}- \mu v+2(u^2+v^2)v-\varepsilon(W_1(x) v+W_2(x)u)=0.\label{Eq:v-intro}
	\end{align} 
	In the limit $\vep=0$ these equations can be considered as a Hamiltonian system with two degrees of freedom.  At  $\vep=0$   this system has a homoclinic orbit corresponding to the bright soliton solution $u(x) =  p_0(x-x_0; \mu) \cos\theta$, $v(x) =  p_0(x-x_0; \mu) \sin\theta$, where $p_0(x; \mu) = \sqrt{\mu}\,\sech (\sqrt{\mu}\,x)$,  and  $x_0$ and $\theta$ are arbitrary reals which, respectively, reflect translational and rotational symmetries of the system with $\vep=0$. The terms proportional to   $\vep$  are considered as   small spatially inhomogeneous perturbations to the Hamiltonian system. To check the persistence of the  homoclinic orbit in the perturbed system, authors of \cite{KCKFB19} construct a two-component Melnikov vector $\vec{M} = [M_1(x_0;  \theta, \mu), M_2(x_0;  \theta, \mu)]$.  Using  formalism from Ref.~\cite{Ya92,Mel03}, the  result is found in  the form
	\begin{align}
	\label{eq:Mel01}
	&M_1(x_0; \theta, \mu) = -\int_{-\infty}^\infty \frac{d W_1(x-x_0)}{dx} p_0^2(x; \mu) dx,\\[2mm]
	\label{eq:Mel02}
	&M_2(x_0; \theta, \mu) = \int_{-\infty}^\infty W_2(x-x_0) p_0^2(x; \mu)dx.
	\end{align} 
	Authors of Ref.~\cite{KCKFB19} argue that the localized stationary state is expected to persist under the perturbation if the entries of the Melnikov vector have a simple zero, i.e., if for some $x_0$, $\theta$ and $\mu$ one has  $ M_1(x_0; \theta, \mu)=M_2(x_0; \theta, \mu)=0$. Then relation  (\ref{Eq:RelMain}) emerges as a  compatibility condition for both entries of $\vec{M}$ to vanish simultaneously: if (\ref{Eq:RelMain}) holds, then  $M_1$ and $M_2$ can be both made zero by adjusting the value of $x_0$.}

\rev{While the Melnikov function is a well-known tool in the study of dynamical system perturbed by a driving force, its  applicability to   system  (\ref{Eq:u-intro})--(\ref{Eq:v-intro}) raises several doubts. First, in Ref.~\cite{Ya92,Mel03} and, apparently, in   most of the literature on the Melnikov theory (e.g.~\cite{Mel01,Mel02,Mel04}) it is assumed that the perturbation is a periodic function of $x$, which is obviously not the case of system (\ref{Eq:u-intro})--(\ref{Eq:v-intro}) with  potential $W(x)$ of generic form. At a less superficial level, we note that the zero of the Melnikov function can be a necessary, but not  sufficient condition for the persistence of the homoclinic orbit under the perturbation. The sufficient conditions  include several more subtle constraints that, in particular, involve the derivatives of the Melnikov vector entries  with respect to their arguments \cite{Ya92}. However, in the case at hand we observe that the  Melnikov vector in (\ref{eq:Mel01})--(\ref{eq:Mel02})   does not depend on $\theta$ (which is a natural consequence of the rotational symmetry). This  suggests that the situation  may be, in some sense, degenerate, and a simple zero of the Melnikov vector may not yet be sufficient. Since these issues are not fully discussed in \cite{KCKFB19}, the results of this paper may    not be fully   rigorous and conclusive, but rather provide an analytical indication at the possible existence of continuous families. }
\rev{To confirm the   predictions of the Melnikov-vector analysis, authors of \cite{KCKFB19} numerically compute families of stationary modes in periodic and quasiperiodic potentials of small, but finite amplitude $\vep$. Authors notice that  the use of different numerical methods yields results of different accuracy, and therefore it is desirable to search for a more efficient method to deepen  the numerical  analysis of found solutions. }

\rev{Motivated by the intriguing outcomes of Ref.~\cite{KCKFB19}, in the present  paper we use a different combination of analytical and numerical approaches to   continuous families of nonlinear modes in  complex potentials that satisfy (\ref{Eq:RelMain}). For brevity, in what follows   we call the potentials that satisfy (\ref{Eq:RelMain}){\it W-dW potentials}.  
	Assuming that a W-dW potential is small, in Sec.~\ref{sec:Asymp} we employ asymptotic methods to construct the  nonlinear modes starting from the limit of zero potential where the stationary solution is readily given in the form of a  bright soliton. We seek for the profile of a nonlinear mode in the form of a power series and show  that the Melnikov-vector conditions (\ref{Eq:RelMain}) enable only the first order of the perturbation theory. For a generic (asymmetric) W-dW potential the asymptotic procedure stops at the second-order terms. To check this prediction,   in Sec.~\ref{Sec:matching} we consider a specific example of asymmetric W-dW  potential. In contrast to Ref.~\cite{KCKFB19}, where sophisticated periodic and quasiperiodic  potentials have been considered, we address the case of a more simple  finite-depth well  potential  which decays exponentially as $x\to \pm \infty$. Using a  transparent numerical shooting method, we confirm that the numerical solutions that bifurcate from the family of bright solitons   satisfy the  equation only  up to $\varepsilon^2$-accuracy. The combination of these results allows to conjecture  that the continuous families in W-dW potentials are in fact  formed by    approximate  solutions that   satisfy the equation only up to $O(\vep^2)$ accuracy.  A similar suggestion has been recently formulated  in \cite{Yang2021}.    We call such approximate   solutions \emph{pseudo-modes}. In Sec.~\ref{sec:Dynamics}  we use numerical simulations to demonstrate that, even though the pseudo-modes do  not correspond to exact stationary modes,  they  play  a distinctive role in the nonlinear dynamics governed by the time-dependent NLSE. Numerical dynamics simulations  show that   for small-amplitude potentials the pseudo-modes  exhibit nearly perfect oscillations of the center of mass around the position  predicted by the asymptotic expansion.  Finally, in Sec.~\ref{sec:SepEps} we extend the numerical shooting method to compute  authentic stationary modes which do not form a continuous family and, in contrast to the solutions of   Ref.~\cite{KCKFB19}, can be found only if the propagation constant $\mu$   is tuned to a certain isolated value.}

\section{Small-amplitude W-dW potential: asymptotic expansions}\label{sec:Asymp}

\rev{In this section, we study the persistence of the continuous family of solitary-wave solutions in a complex potential of small amplitude described by the formal parameter $\vep\ll 1$. In contrast to the analysis of Ref.~\cite{KCKFB19}, where the Melnikov theory has been employed for this purpose, we take  a somewhat blunter  approach and try to construct asymptotic power series that  directly describe the deformation of the continuous family as the amplitude of the complex potential increases departing from zero. Unsurprisingly,  for a complex potential of general form the asymptotic procedure  terminates already in the first order of amplitude. However, if the complex potential if of W-dW form, then the the asymptotic procedure can proceed   beyond the first  order. Therefore, the W-dW relations (\ref{Eq:RelMain}) appear as a necessary  condition for the persistence of the continuous family under the perturbation by a non-$\PT$-symmetric small-amplitude complex potential. However, even if the potential is of  W-dW-type, the asymptotic procedure  generically terminates at the second  order. This outcome leads us to a conjecture that   the W-dW relation  \emph{per~se} is   not sufficient for the persistence of the continuous family.}

\rev{For $\varepsilon=0$ system (\ref{Eq:u-intro})-(\ref{Eq:v-intro})  has a well-known family of   bright soliton solutions  
\begin{gather*}
u_0(x)= \sqrt{\mu}\,\sech(\sqrt{\mu}\,(x-x_0))\cos \theta,\quad v_0(x)=\sqrt{\mu}\,\sech(\sqrt{\mu}\,(x-x_0))\sin \theta,
\end{gather*}
where $x_0$ and $\theta$ are arbitrary reals. Due to the rotational symmetry of the NLSE, without loss of generality one can fix $\theta=0$. We therefore set
\begin{gather*}
u_0(x)= \sqrt{\mu}\,\sech(\sqrt{\mu}\,(x-x_0)),\quad v_0(x)=0.
\end{gather*}
Assume that for nonzero, but fixed $\vep \ll 1$, system (\ref{Eq:u-intro})--(\ref{Eq:v-intro}) has a family of solitary wave solution, i.e., there exits functions $u(x; \mu)$ and $v(x; \mu)$, where $\mu$ changes continuously inside some interval.  Our strategy in this section is to approach solutions $u(x; \mu)$ and $v(x; \mu)$ from the limit $\vep=0$. We therefore fix $\mu>0$ and 
seek for  solutions of (\ref{Eq:u-intro})-(\ref{Eq:v-intro})  in the form of asymptotic expansions
\begin{align*}
&u(x)=u_0(x)+\varepsilon u_1(x)+\varepsilon^2 u_2(x)+\ldots,\\[2mm]
&v(x)=\phantom{u_0(x0)+}\varepsilon v_1(x)+\varepsilon^2 v_2(x)+\ldots.
\end{align*}
(Since $\mu$ is fixed in our computations, hereafter we do not indicate  the dependence on $\mu$   explicit  and write $u(x)$ instead of $u(x;\mu)$,  $v(x)$ instead of $v(x;\mu)$,  etc.)}

Balance of the terms of the order $O(\varepsilon)$ yields
\begin{align}
&{\cal L}_6u_{1}=W_1(x)u_0(x),\label{Eq:L6_01}\\[2mm]
&{\cal L}_2v_{1}=W_2(x)u_0(x),\label{Eq:L2_01}
\end{align}
where we have introduced operators
\begin{align*}
&{\cal L}_n:=\frac{d^2\,}{dx^2}-\rev{\mu}+ n \rev{u_0^2(x)}, \quad n\in \{2, 6\}. 
\end{align*}
Since 
${\cal L}_6 u_{0,x}(x) =0$,  the operator  ${\cal L}_6$   has nonempty kernel. 
This implies that Eq.~(\ref{Eq:L6_01}) has a solitary-wave solution if the orthogonality condition holds:
\begin{gather}
\int_{-\infty}^\infty W_1(x)u_0(x)u_{0,x}(x)~dx=-\frac{1}{2}\int_{-\infty}^\infty W_{1,x}(x)u_0^2(x)~dx=0.\label{Eq:Cond01}
\end{gather}
The kernel of operator  ${\cal L}_2$ is also nonempty, because 
${\cal L}_2 u_0=0$.
Therefore Eq.~(\ref{Eq:L2_01}) has a solitary wave solution if
\begin{gather}
\int_{-\infty}^\infty W_2(x)u_0^2(x)~dx=0.\label{Eq:Cond02}
\end{gather}
Therefore, two nontrivial conditions [Eqs.  (\ref{Eq:Cond01}) and (\ref{Eq:Cond02})] emerge already  in the first order of the asymptotic theory. \rev{They cannot be satisfied for a complex potential of general shape.} However, if \rev{the potential is of W-dW form, i.e.,}
\begin{gather}
W_{1,x}(x)=CW_2(x),\label{Eq:CondMain}
\end{gather}
where $C$ is an arbitrary real, then the   conditions   (\ref{Eq:Cond01}) and (\ref{Eq:Cond02})
coincide. This agrees completely  with the result of \cite{KCKFB19}. In this case  the  admissible values of $x_0$ are determined at the first step of the asymptotic procedure by the equation
\begin{gather}
\int_{-\infty}^\infty  {W_2(x)}{\sech^2(\rev{\sqrt{\mu}}(x-x_0))}~dx=0.\label{Eq:Cond_x}
\end{gather}
If the solvability conditions  of (\ref{Eq:Cond01})-(\ref{Eq:Cond02})  are fulfilled, then the general solitary-wave solutions of   system (\ref{Eq:Cond01})--(\ref{Eq:Cond02}) have  the form
\begin{align}
&u_1(x)=\tilde u_{\rev{1}}(x)+C_1u_{0,x}(x),\label{Eq:u_c1}\\[2mm]
&v_1(x)=\tilde v_{\rev{1}}(x)+C_2u_{0}(x),\label{Eq:v_c2}
\end{align}
where $C_{1,2}\in \mathbb{R}$  are arbitrary constants and $\tilde u_{\rev{1}}(x)$ and $\tilde v_{\rev{1}}(x)$ are some fixed solitary wave solutions of (\ref{Eq:Cond01}) and (\ref{Eq:Cond02}), respectively. Thus  the functions $u_1(x)$ and  $v_1(x)$ are not yet uniquely defined. 

In order to specify the constants $C_{1,2}$ let us analyze the terms of order $O(\varepsilon^2)$. Balance of the terms yields the system
\begin{align}
&{\cal L}_6u_2=-6u_0u_1^2-2u_0v_1^2+W_1u_1-W_2v_1,
\label{Eq:L06_02}\\[2mm]
&{\cal L}_2v_2=-4u_0u_1v_1+W_1v_1+W_2u_1,
\label{Eq:L02_02}
\end{align}
(we simplify the notations taking $u_{k}(x):=u_{k}$,  $k=0,1,2$, $v_{k}(x):=v_{k}$, $W_k(x):=W_k$, $k=1,2$ and $W_{k,x}(x):=W_{k,x}$, $k=1,2$). The solvability conditions for (\ref{Eq:L06_02})-(\ref{Eq:L02_02}) are
\begin{align}
&\int_{-\infty}^\infty (-6u_0u_1^2-2u_0v_1^2+W_1u_1-W_2v_1
)  u_{0,x}~dx=0,\label{Eq:IntCond01}\\[2mm]
&\int_{-\infty}^\infty (-4u_0u_1v_1+W_1v_1+W_2u_1
)  u_0~dx=0. \label{Eq:IntCond02}
\end{align}
These conditions should be satisfied by the proper choice of constants $C_{1,2}$ in (\ref{Eq:u_c1})-(\ref{Eq:v_c2}). 
Substituting (\ref{Eq:u_c1})-(\ref{Eq:v_c2}) into (\ref{Eq:IntCond01})-(\ref{Eq:IntCond02}) and collecting the terms with $C_1$ and $C_2$ separately, one arrives at the system of linear equations
\begin{align}
&A_{11}C_1+A_{12}C_2+F_1=0,\label{Eq:LinSys01}\\[2mm]
&A_{21}C_1+A_{22}C_2+F_2=0.\label{Eq:LinSys02}
\end{align}
Taking into account condition (\ref{Eq:CondMain}), it is straightforward to show that (see Appendix~\ref{Sect:Coeff} for detailed calculations)
\begin{align*}
&A_{12}=A_{22}=0,\\[2mm]
&A_{21}=8\int_{-\infty}^\infty u_0^2u_{0,x}\tilde{v}_1~dx,\quad A_{11}=-\frac{C}2A_{12},
\end{align*}
\begin{align*}
&F_1=C\int_{-\infty}^\infty W_2 u_0\tilde{u}_1~dx+\int_{-\infty}^\infty\tilde{v}_1(2W_2u_{0,x}+W_{2,x}u_0)~dx,
\\[2mm]
&F_2=-2\int_{-\infty}^\infty W_2\tilde{u}_1u_0~dx.
\end{align*}
Therefore $C_2$ in fact  does not enter   equations (\ref{Eq:LinSys01}) and (\ref{Eq:LinSys02}). This means that the system (\ref{Eq:LinSys01})-(\ref{Eq:LinSys02}) has a solution if the following condition holds: 
\begin{gather*}
I:=F_2 C+2F_1=2\int_{-\infty}^\infty\tilde{v}_1(2W_2u_{0,x}+W_{2,x}u_0)~dx=0.
\end{gather*}
Generically, $I\ne 0$ and the asymptotic procedure terminates.  \rev{One can try to make $I$ equal to zero by adjusting the value of the parameter $\mu$. However, even if the condition $I=0$ is satisfied for some isolated values of $\mu$, this still contradicts to the assumption of existence of a continuous family. Moreover,} even if $I=0$, the procedure is still not self-consistent, because  $C_2$  cannot be determined unambiguously. 

\rev{Notice that if the potential is $\PT$ symmetric, i.e., $W_1(x)=W_1(-x)$ and $W_2(x)=-W_2(-x)$, then, due to the parity of the solutions, the solvability conditions are automatically satisfied either at $\vep$- and $\vep^2$-order by setting $x_0=0$.}

The upshot of   our analysis is that for a generic (i.e., non-$\PT$-symmetric) small-amplitude W-dW potential  the asymptotic procedure allows to construct  an approximation that satisfies the stationary equation (\ref{Eq:phi}) with $O(\varepsilon^2)$-accuracy.  However, the  procedure fails to produce a more exact result. Strictly speaking, this may be due to the prescribed analytic form of expansion (power series with respect to $\varepsilon$) that may be not appropriate for the stationary  solution. Therefore in Sec.~\ref{Sec:matching} we employ another approach for the problem.
 
\section{W-${\bf {\rm d}}$W  well of finite depth: a numerical study}\label{Sec:matching}

\rev{In this section, we support the results of our asymptotic analysis with a numerical study of a W-dW potential of finite amplitude. In contrast to the analysis of Ref.~\cite{KCKFB19}, where the Levenberg-Marquardt algorithm and Matlab boundary-value solver have been employed for  numerical search of stationary states, here we use a shooting-type approach. Various modifications of this   method  have been previously applied for real \cite{shoot01,AZ07}, dissipative \cite{shoot02}, $\PT$-symmetric \cite{shoot03} and Wadati \cite{KZ14} potentials. Advantages  of this method consist in its transparency and geometric visualization, because the stationary modes can be searched as an intersections of certain two-dimensional curves.   We argue that the system of equations that determines a continuously-differentiable nonlinear mode is, generically speaking, overdetermined (i.e., the number of equations in the system is larger than the number of unknowns).  Therefore it is generically impossible to find an exact solution to this system, and only approximate solutions with nonzero residual  are possible. We call such nonzero-residual solutions \emph{pseudo-modes}, to distinguish  them from    the authentic  continuously-differentiable   modes.  Varying the amplitude of the W-dW potential, we numerically confirm that  for the pseudo-modes of the simplest form the residual behaves as $O(\vep^2)$. This outcome confirms the finding of the asymptotic analysis in the previous section. There also exist pseudomodes of more complex shapes which have not been captured by our asymptotic procedure. For these pseudomodes  the residual is also generically  different from  zero, but depends on   $\vep$ according to a more complex law.}

Consider now Eq.~(\ref{Eq:phi}) with a potential that is a {\it W-dW well of finite depth}
\begin{gather*}
\lim_{x\to-\infty}W(x) = \lim_{x\to\infty}W(x)=\lim_{x\to-\infty}W_x(x) = \lim_{x\to\infty}W_x(x) =   0. 
\end{gather*}
We also assume that $W(x)$ and its derivative   decay     {\it exponentially} when $x\to\pm\infty$. A prototypical example is $W(x)=W_1(x)+iW_2(x)$, where
\begin{gather}
W_1(x)=-\frac{A}{e^{\alpha x}+Be^{-\beta x}},\quad W_2(x)=W_{1,x}(x),\quad   \alpha,\beta,A, B>0. \label{Eq:W12}
\end{gather}
\rev{This   shape of $W_1(x)$ guarantees that, for generic values of parameters $ \alpha,\beta,A, B$,  the  resulting complex  W-dW potential is  neither ${\cal PT}$ symmetric nor Wadati-type.} Its   real part $W_1(x)$ has   the unique local minimum situated at $x=(\alpha+\beta)^{-1}\ln(\alpha^{-1} \beta B)$.

\rev{Fix $\mu>0$.}
Let $S^+$  be the class of solutions for  Eq.~(\ref{Eq:phi}) that tend to zero when $x\to+\infty$, i.e., 
\begin{gather*}
S^+=\{\phi(x)|\, \phi(x)\to 0,\quad x\to+\infty\}.
\end{gather*}
Then $\phi(x)\in S^+$ has the asymptotic behavior
\begin{gather}
\phi(x)=e^{-\rev{\sqrt{\mu}}x}(C^++ o(1)),\quad x\to +\infty,\label{Eq:As+}
\end{gather}
where $C^+$ is a complex constant. {We assume that any $\phi(x)\in S^+$ uniquely defines $C^+$  in the asymptotic relation (\ref{Eq:As+}) and \emph{vice versa}, for any $C^+$ there exists the unique  $\phi(x)\in S^+$ which obeys (\ref{Eq:As+})   (for real potentials the existence of this one-to-one correspondence  was proven in \cite{AZ07}).}
Note that if $\phi(x)\in S^+$ with constant $C^+=|C^+|e^{i\theta^+}$  in asymptotic relation (\ref{Eq:As+}), then the phase-rotated solution $\phi(x)e^{-i\theta^+}\in S^+$  is associated with real constant $|C^+|$ in  (\ref{Eq:As+}).  Similarly,  let   $S^-$  be the class of solutions for Eq.~(\ref{Eq:phi}) that tend to zero when $x\to-\infty$, i.e., 
\begin{gather*}
S^-=\{\phi(x)|\,\phi(x)\to 0,\quad x\to-\infty\}.
\end{gather*}
Then $\phi(x)\in S^-$ has the asymptotic behavior
\begin{gather}
\phi(x)=e^{\rev{\sqrt{\mu}}x}(C^-+ o(1)),\quad x\to -\infty.\label{Eq:As-}
\end{gather}
and  any $\phi(x)\in S^-$ with complex constant $C^-=|C^-|e^{i\theta^-}$ can be phase-rotated such that   $\phi(x)e^{-i\theta^-}\in S^-$ corresponds to  real constant $|C^-|$ in (\ref{Eq:As-}).

If $\phi(x)$ is a localized   solution for  Eq.~(\ref{Eq:phi}), then $\phi(x)\in S^+\cap S^-$. The constants $C^+$ and $C^-$ that uniquely define the behavior of   $\phi(x)$ at $x\to \pm \infty$ are generically complex. Since the   solution $\phi(x)$ is physically indistinguishable from its phase-rotated counterpart $\phi(x)e^{-i\theta}$, we can assume that   one of the constants (either $C^+$ or $C^-$) is real. However, the second constant is generically complex. 

Consider solutions of Eq.~(\ref{Eq:phi})  on semiaxes, $\mathbb{R}^+$ and $\mathbb{R}^-$. Let a solution $\phi^+(x)\in S^+$ be defined on $\mathbb{R}^+$ 
having {\it real} constant $C^+$ in (\ref{Eq:As+}). Also, let a solution $\phi^-(x)\in S^-$ be defined on $\mathbb{R}^-$ 
with {\it real} constant $C^-$ in (\ref{Eq:As-}).
In order to get a solution that is continuously differentiable  on the entire axis $\mathbb{R}$, one has to find two phases $\theta^+$ and $\theta^-$ such that the matching conditions hold
\begin{align*}
e^{i\theta^-}\phi^-(0)&=e^{i\theta^+}\phi^+(0),\\[2mm]
e^{i\theta^-}\phi^-_x(0)&=e^{i\theta^+}\phi^+_x(0),
\end{align*}
or, alternatively
\begin{align}
\phi^-(0)&=e^{i\theta}\phi^+(0),\label{Eq01}\\[2mm]
\phi^-_x(0)&=e^{i\theta}\phi^+_x(0),\label{Eq02}
\end{align}
where $\theta=\theta^+-\theta^-$.
We note that the system (\ref{Eq01})-(\ref{Eq02}) implies that
\begin{align}
&|\phi^-(0)|=|\phi^+(0)|,\label{Eq_modF}\\[2mm]
&|\phi^-_x(0)|=|\phi^+_x(0)|,\label{Eq_modDer}\\[2mm]
&{\rm arg}~\phi^-(0)={\rm arg}~\phi^+(0)+\theta,\label{Eq_phaseF}\\[2mm]
&{\rm arg}~\phi^-_x(0)={\rm arg}~\phi^+_x(0)+\theta.\label{Eq_phaseDer}
\end{align}
This is a system of {\it four} real equations that includes only {\it three} unknowns $C^+$, $C^-$ and $\theta$. Generically, it does not have solutions.
It might have solutions in the presence of some additional symmetries or integrals (\rev{in particular, in the $\PT$-symmetric case the shooting approach can be reduced to solution of only one equation with respect to one real unknown \cite{shoot03}, while for Wadati potential the situation reduces to a system of three equations with respect to three real unknowns \cite{KZ14}}). However,   for a generic complex potential the existence of localized solutions for Eq.~(\ref{Eq:phi}) is dubious.

In order to check whether the system (\ref{Eq_modF})-(\ref{Eq_phaseDer}) has a solution for a given W-dW potential we use  the following strategy.\medskip

{\bf 1.} {  For fixed $\varepsilon$ \rev{and $\mu$}, compute the values of $C^\pm\in\mathbb{R}$ such that the equations (\ref{Eq_modF})-(\ref{Eq_modDer}) hold.} Algorithmically this was done as follows.
Denote $|\phi^-(0)|=R^-$, $|\phi^-_x(0)|=r^-$, $|\phi^+(0)|=R^+$, $|\phi^+_x(0)|=r^+$.  In view of (\ref{Eq:As+}) and (\ref{Eq:As-}),  $R^-\equiv R^-(C^-)$, $r^-\equiv r^-(C^-)$, $R^+\equiv R^+(C^+)$, $r^+\equiv r^+(C^+)$. Plot on the plane $(R,r)$ two curves:
$\gamma^-=\{(R^-(C^-),r^-(C^-))| C^-\in(0;C^-_{\max})\}$, parametrized by $C^-$ and $\gamma^+=\{(R^+(C^+),r^+(C^+))|  C^+\in(0;C^+_{\max})\}$ parametrized by $C^+$. At the point of intersection of these curves $R^-(C^-)=R^+(C^+)$ and $r^-(C^-)=r^+(C^+)$. If this point is determined, the values of $C^+$ and $C^-$ are found such that (\ref{Eq_modF})-(\ref{Eq_modDer}) are satisfied. The procedure involves computation of $\phi^\pm(0)$ \rev{and $\phi_x^\pm(0)$} by given $C^\pm$. This can be done by standard Runge-Kutta method that solves ODE  (\ref{Eq:phi}) with initial conditions
\begin{gather*}
\phi^+(x_+)=C^+e^{- \rev{\sqrt{\mu}}x_+}, \quad \phi^+_x(x_+)=-\rev{\sqrt{\mu}}C^+e^{- \rev{\sqrt{\mu}}x_+}
\end{gather*}
for $\phi^+(x)$, and 
\begin{gather*}
\phi^-(x_-)=C^-e^{ \rev{\sqrt{\mu}}x_-}, \quad \phi^-_x(x_-)=\rev{\sqrt{\mu}}C^-e^{ \rev{\sqrt{\mu}}x_-}
\end{gather*}
for $\phi^-(x)$. The value $x_+>0$ has to be chosen large enough in such a way that \rev{the correction $o(1)$ can be neglected safely in (\ref{Eq:As+}). 
Similarly,  the value $x_-<0$ has to be chosen large negative such that $o(1)$ can be neglected in  (\ref{Eq:As-}).   If  $x_+$ and $x_-$ are chosen properly, then the further increase (respectively, decrease) of $x_+$ (respectively, $x_-$) does not affect   the values   of $C^+$ and $C^-$ corresponding to the intersection point.   }\medskip

{\bf 2.} {  Having $\phi^+(0)$, $\phi^-(0)$, $\phi^+_x(0)$, $\phi^-_x(0)$ that correspond to the intersection  $\gamma^+\cap\gamma^-$,
	compute the values
\begin{align}
\theta&= -{\rm arg}~\phi^+(0) + {\rm arg}~\phi^-(0),\label{Theta_1}\\[2mm]
\tilde{\theta}&=-{\rm arg}~\phi^+_x(0) + {\rm arg}~\phi^-_x(0).\label{Theta_2}
\end{align}
}
The condition for solvability of (\ref{Eq_modF})-(\ref{Eq_phaseDer}) is
\begin{gather}\delta\equiv \theta-\tilde\theta =0.\label{FinalEq}
\end{gather}
If this condition is satisfied, then the piecewise-defined  function 
\begin{gather}
\label{eq:pseudo}
\phi(x)=\left\{
\begin{array}{cc}
\phi^-(x)e^{-i\theta},& x\leq 0\\[2mm]
\phi^+(x),& x\geq 0
\end{array}
\right. .
\end{gather}
solves all four equations of system  (\ref{Eq_modF})--(\ref{Eq_phaseDer}) and therefore corresponds to an authentic stationary mode which is continuously differentiable.  However,  if $\delta \ne 0$, then function  (\ref{eq:pseudo}) solves only three of four equations. In what follows, we will say that such a function with $\delta\ne 0$  corresponds to a  \emph{pseudo-mode}.

\begin{figure}
	\begin{center}
		\includegraphics[width=0.8\textwidth]{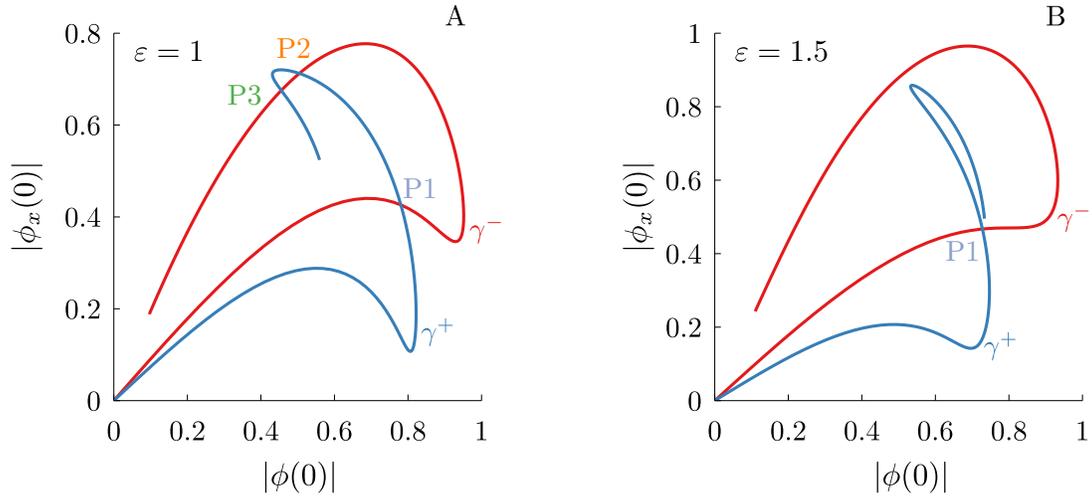}%
	\end{center}
	\caption{The curves $\gamma^+$ (blue) and $\gamma^-$ (red) and their intersections (labeled as P1, P2, P3) for $\varepsilon=1$ (panel A) and $\varepsilon=1.5$ (panel B).  For all curves, $C^\pm\in [0, 40]$. \rev{Here $\mu=1$.}}
	\label{fig:Intersect}
\end{figure}

\begin{figure}
	\begin{center}
		\includegraphics[width=0.81\textwidth]{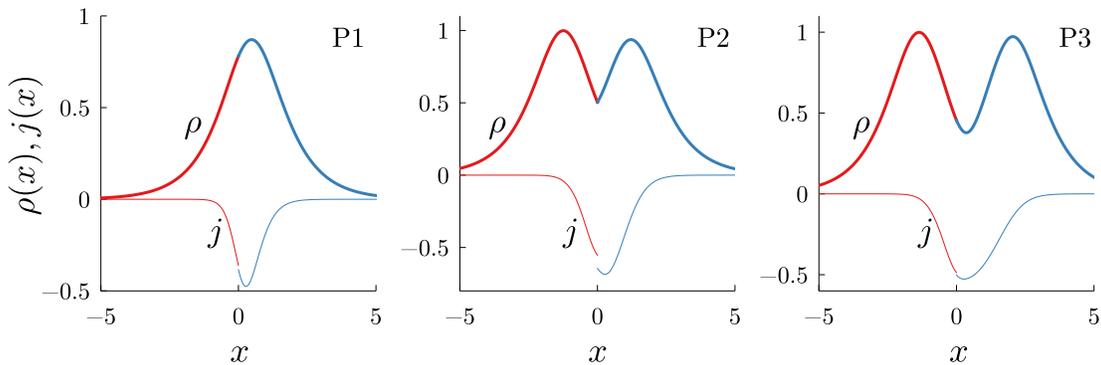}%
	\end{center}
	\caption{Shapes of the pseudo-modes corresponding to the intersection points P1, P2, P3 in Fig.~\ref{fig:Intersect}. }
	\label{fig:Pseudo}
\end{figure}

 \begin{figure}
	\begin{center}
		\includegraphics[width=0.8\textwidth]{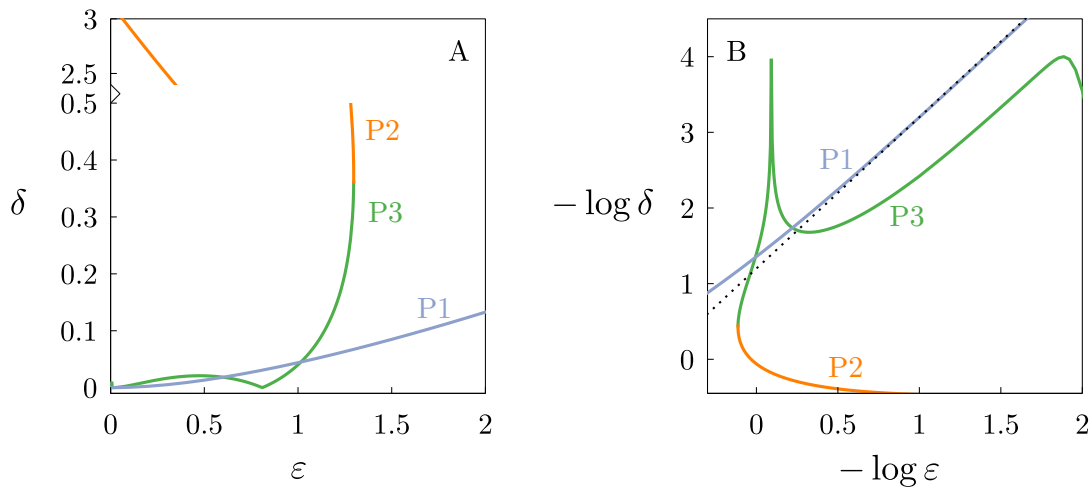}%
	\end{center}
	\caption{Dependencies of $\delta$ plotted with  linear scale (A) and with  log-log scale (B). For the reference, in (B) we also plot  the straight dotted line with the slope equal to two. In the linear scale, this line corresponds to $\delta\propto \varepsilon^2$. Notice that the vertical axis  is broken in (A).  Labels P1, P2, P3 correspond to the intersection points in Fig.~\ref{fig:Intersect}. \rev{Here $\mu=1$.}}
	\label{fig:Th_vs_eps}
\end{figure}

As an example, we chose a nonsymmetric W-dW potential of the class (\ref{Eq:W12}) having
\begin{gather}
W_1(x)=-\frac1{e^x+e^{-3x}},\quad W_2(x)=\frac{e^x-3e^{-3x}}{(e^x+e^{-3x})^2}.\label{Eq:WdW}
\end{gather}
Figure~\ref{fig:Intersect} presents  two plots of    curves $\gamma^\pm$ computed for \rev{$\mu=1$ and } two different values of $\vep$. We observe that the curves can have multiple intersection points, and for different $\vep$ the shapes of the curves can be significantly different, i.e.,  the intersection points   can emerge or disappear as $\vep$  changes. Here we focus on three first intersection points that are labeled as P1, P2, P3 in Fig.~\ref{fig:Intersect}(A). Notice that with the increase of $\vep$  points P2 and P3 merge and then   disappear  as illustrated in Fig.~\ref{fig:Intersect}(B).  To visualize the pseudo-modes   that correspond to the chosen intersection points, we introduce real-valued piecewise-defined functions 
\begin{gather*}
\rho(x)=\left\{
\begin{array}{cc}
|\phi^-(x)|,& x\leq 0,\\[2mm]
|\phi^+(x)|,& x\geq 0,
\end{array}
\right. \quad j(x)=\left\{
\begin{array}{cc}
i(\overline{\phi^-_x} \phi^-  - \overline{\phi^-} \phi^-_x),& x\leq 0,\\[2mm]
i(\overline{\phi^+_x} \phi^+  - \overline{\phi^+} \phi^+_x),& x\geq 0.
\end{array}
\right.
\end{gather*}
These functions do not  depend  on the rotation $\theta$ and   are therefore    especially convenient. Hereafter the overline means complex conjugation. Notice that in the physical context function $j(x)$ can be interpreted as energy flux across the (pseudo)-mode. By construction, for each pseudo-mode   $\rho(x)$ is  continuous, but it is not necessarily smooth;  function $j(x)$ must be continuous for authentic stationary modes, but may have a discontinuity at $x=0$ for pseudo-modes.  

Figure~\ref{fig:Pseudo} presents the  pseudo-modes   corresponding  to intersections P1, P2, and P3.  We observe that for each shown pseudo-mode   the corresponding function $j(x)$  has a jump at $x=0$. Additionally, for P2 the cusp of $\rho(x)$ is well-visible at $x=0$. The pseudo-mode  at   the first intersection point  P1  resembles the bright soliton, i.e., corresponds to the approximate solution constructed above in Sec.~\ref{sec:Asymp} by means of the asymptotic expansions. Solutions at the next intersections P2 and P3 have more sophisticated shapes and therefore cannot    be captured by the asymptotic expansions   developed above. 

Discontinuous  shapes of pseudo-modes  plotted in Fig.~\ref{fig:Pseudo} suggest that those solutions     do not correspond to authentic   continuously differentiable stationary  modes. Indeed, evaluating    the    solvability indicator   $\delta$, we observe that    it is generically different from zero. Dependencies   $\delta$ on $\vep$ are presented in Fig.~\ref{fig:Th_vs_eps} in  linear and log-log scales. For the simple pseudo-mode corresponding to the first intersection point P1 we observe that   the dependence of $\log \delta$ on $\log \varepsilon$
is well approximated by linear function with the slope close to 2. This again agrees with the above asymptotic analysis and  suggests that this pseudo-mode 
solves 
Eq.~(\ref{Eq:phi})  for all $x$ \emph{except for} $x=0$, where the derivative of    function $\phi(x)$ has  a jump  that is of order $O(\varepsilon^2)$.

For the pseudo-modes corresponding to P2 and P3, the  dependencies $\delta(\epsilon)$ are more sophisticated and cannot be described by   a simple quadratic law. In the meantime, it is remarkable, that for $\vep\approx 0.8$ the $\delta(\vep)$-function   corresponding to P3     \rev{has a  zero   [in the log-log plot in Fig.~\ref{fig:Th_vs_eps}(B) it corresponds to a spike]}. This suggests that for some isolated value of $\vep$  close to 0.8 an authentic continuously differentiable solution can potentially be found. Solutions of this type will be discussed below in Sec.~\ref{sec:SepEps}.   Summarizing the analysis of the present   section,  we have to conclude that  for an arbitrarily chosen value of $\varepsilon$,  Eq.~(\ref{Eq:phi}) with potential (\ref{Eq:WdW}) only admits pseudo-modes and    no  stationary   modes.

\section{Nonlinear dynamics in W-$\rm d$W potentials}\label{sec:Dynamics}

\rev{The goal of this section is to elucidate nonlinear dynamics governed by the time-dependent equation (\ref{Eq:Phi}) with a W-dW potential.
One of important analytical results of this section consists in the approximate conservation law that constraints the nonlinear dynamics in W-dW potentials.   A dynamical invariant of motion for solitons in W-dW potentials has been earlier reported on in \cite{Kominis_OC15,Kominis_PRA15}. Our result differs from the previous in two important aspects. First,     the dynamical invariant from   \cite{Kominis_OC15,Kominis_PRA15} is obtained with a qualitative approach which treats a soliton as a particle with some mass, velocity and position. However, in the framework of the time-dependent Eq.~(\ref{Eq:Phi}), some of the  effective quantities (namely, the soliton position and velocity) are not well-defined. Our conservation law is obtained in terms of the wavefunction $\Phi$. It   does not rely on  the effective-particle formalism and is therefore valid for localized nonlinear waves of arbitrary shape (not necessarily  single-soliton-shaped). Second, our conservation law is only \emph{approximate}. Our dynamical invariant is exactly time-independent  only in the limit of zero   amplitude of W-dW potential, while  for nonzero $\vep$  the temporal derivative of our invariant is of the $\vep^2$-order. In this section we also perform numerical dynamical runs of the time-dependent equation~(\ref{Eq:Phi}) and observe that even though the   pseudo-modes do not correspond to authentic stationary states, they feature meaningful  nonlinear dynamics  associated with the persistent oscillations of the    soliton center  around the position $x_0$ obtained from the above asymptotic analysis. }

\subsection{Approximate conservation law and necessary steady-state conditions}\label{sec:ConsLaws}

A natural question emerges on   whether the pseudo-modes encountered in the previous sections  have   any signature  in the nonlinear time-dependent dynamics governed by the non-stationary equation (\ref{Eq:Phi}).  This issue will be addressed in the present section.  
However, let us first outline some general features of nonlinear dynamics in W-dW potentials.  Let  $\Phi(x,t)$ be a localized wavepacket  whose dynamics is governed by   equation  (\ref{Eq:Phi}).  We introduce the squared $L^2$-norm of the solution  (in the optical context it can be interpreted as the beam power) and the location of the center of the wavepacket: 
\begin{equation}
N(t) = \int_{-\infty}^{\infty}|\Phi|^2dx, \quad X(t) = N^{-1}(t)  \int_{-\infty}^{\infty}x|\Phi|^2dx.
\end{equation}
Computing the temporal derivative of $N(t)$ we  obtain  the standard ``balance equation''
\begin{equation}
\label{eq:Nt}
N_t = 2\varepsilon \int_{-\infty}^{\infty}   W_2|\Phi|^2dx.
\end{equation}
For a shape-preserving stationary mode $\Phi = e^{i\mu t} \phi(x)$ this gives an obvious condition 
 \begin{equation}
\label{eq:cond2}
2\varepsilon  \int_{-\infty}^\infty W_2  |\phi|^2dx = -\frac{2\varepsilon}{C} \int_{-\infty}^\infty W_1 \left(\frac{d\,}{dx}|\phi|^2\right)  dx = 0. 
\end{equation}
This condition generalizes   that     derived above in the first-order perturbation theory [see  Eq.~(\ref{Eq:Cond02})].

Additionally,  introducing the momentum  $P(t) = i\int_{-\infty}^{\infty}(\overline{\Phi_x}\Phi - \Phi_x \overline{\Phi})dx$, from Eq.~(\ref{Eq:Phi}) we compute
\begin{equation}
\label{eq:pt}
P_t = -2C\varepsilon \int_{-\infty}^{\infty} W_2|\Phi|^2 dx  + 2i \varepsilon \int_{-\infty}^{\infty} W_2(\Phi_x^*\Phi - \Phi_x\Phi^*) dx.
\end{equation}
An additional calculation yields  
\begin{eqnarray}
\varepsilon \frac{d\,}{dt} \int_{-\infty}^{\infty} W_1 |\Phi|^2dx = i\varepsilon C\int_{-\infty}^{\infty} W_2(\Phi_x^*\Phi - \Phi_x\Phi^*)dx 
+2\varepsilon^2 \int_{-\infty}^{\infty} W_1 W_2|\Phi|^2dx.
\end{eqnarray}
Combining the latter  relations with (\ref{eq:Nt}) and  (\ref{eq:pt}), we obtain
\begin{equation}
\label{eq:law}
\frac{d\,}{dt}\left(P + C N - \frac{2\varepsilon}{C}\int_{-\infty}^{\infty} W_1 |\Phi|^2dx \right) = -\frac{4\varepsilon^2}{C} \int_{-\infty}^{\infty} W_1 W_2 |\Phi|^2dx.
\end{equation}
For small $\varepsilon$,  the latter equality   can be considered as an ``approximate'' conservation law which is specific to small-amplitude W-dW potentials. \rev{In the line with   findings of Sec.~\ref{sec:Asymp}, this result indicates that a careful attention to   the $\vep^2$-order-behavior  is crucial for the most precise description of nonlinear waves in W-dW potentials.}

For a   stationary mode, the left-hand side of (\ref{eq:law}) is zero, which leads to another  necessary condition  for the shape of the solitary state:
\begin{equation}
\label{eq:cond}
-\frac{4\varepsilon^2}{C} \int_{-\infty}^{\infty} W_1 W_2 |\phi|^2dx = \frac{2\varepsilon^2}{C^2} \int_{-\infty}^{\infty} W_1^2 \left(\frac{d\,}{dx}|\phi|^2\right)  dx = 0. 
\end{equation}
Introducing the transverse current $j(x)$ across the stationary state  
\begin{equation}
j(x)  = i(\overline{\phi_x}\phi - \overline{\phi}\phi_x),
\end{equation}
we obtain the standard result which interrelates the derivative of the current and the gain-and-loss distribution:
\begin{equation}
j_x = 2\varepsilon W_2 |\phi|^2.
\end{equation} 
More interestingly,    for W-dW potentials  we obtain 
\begin{equation}
C\int_{-\infty}^{\infty} j(x) W_2(x)dx  + 2\varepsilon \int_{-\infty}^{\infty} W_1W_2|\phi|^2 dx = 0.
\end{equation}
Therefore, condition (\ref{eq:cond}) is equivalent to  
\begin{equation}
\label{eq:cond3}
\int_{-\infty}^{\infty} j(x) W_2(x)dx = 0.
\end{equation} 
Comparing  (\ref{eq:cond2}) and   (\ref{eq:cond3}), we observe that for a stationary mode the imaginary part of the potential $W_2$ must be orthogonal not only to the squared modulus of the wavefunction but also to the  shape the transverse current distribution. 

\subsection{Numerical simulations of nonlinear dynamics}\label{sec:Simulat}

\begin{figure}
	\begin{center}
	 	\includegraphics[scale=0.9]{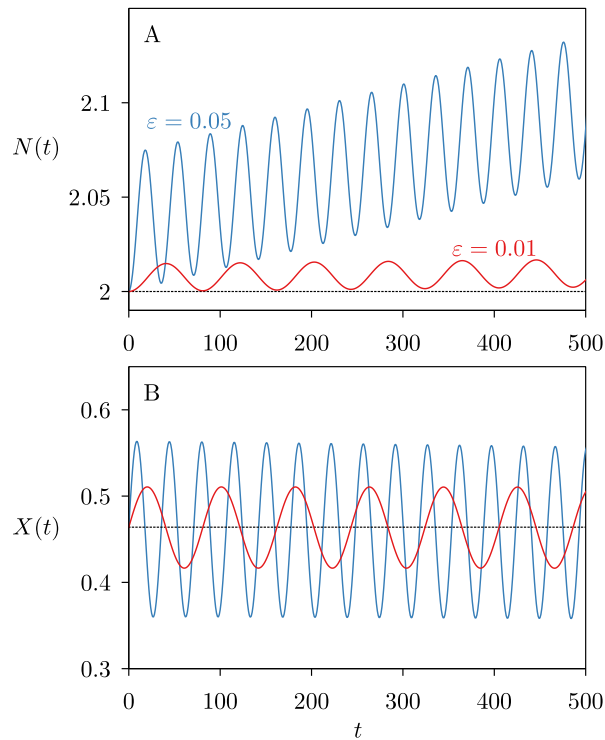}%
	\end{center}
	\caption{Results of the nonlinear dynamics simulations for $\vep=0.01$ (red color) and $\vep=0.05$ (blue color)   for the initial condition $\Phi(x, 0)  = \sech(x-x_0)$, where $x_0$ is a solution of (\ref{Eq:Cond02}).}
	\label{fig:dyn01}
\end{figure}

Let us now turn to dynamics of  the pseudo-mode solitary waves that satisfy  Eq.~(\ref{Eq:phi}) with  $\varepsilon^2$-accuracty (see  Sec.~\ref{sec:Asymp}). As a model example, we again choose the W-dW potential (\ref{Eq:W12}). First, we solve the Cauchy problem with the initial condition $\Phi(x, 0)  = \sech(x-x_0)$, where $x_0$ is chosen to satisfy the compatibility condition (\ref{Eq:Cond02}) that emerges in   the first order  of the perturbation procedure. Numerical solution of Eq.~(\ref{Eq:Cond02}) gives $x_0\approx 0.4640$.   Representative examples of our dynamical simulations are shown in Fig.~\ref{fig:dyn01} for the squared norm $N(t)$ and center of mass $X(t)$.  For sufficiently small $\varepsilon$ we observe that the plotted characteristics  feature small-amplitude nearly periodic oscillations. For small $\varepsilon$ the periodicity  is almost perfect, whereas  for larger $\varepsilon$ a    slow drift appears. Amplitude of the oscillations and the  drift velocity   naturally become stronger  with the increase of amplitude of the potential $\varepsilon$.

\begin{figure}
	\begin{center}
		\includegraphics[scale=0.9]{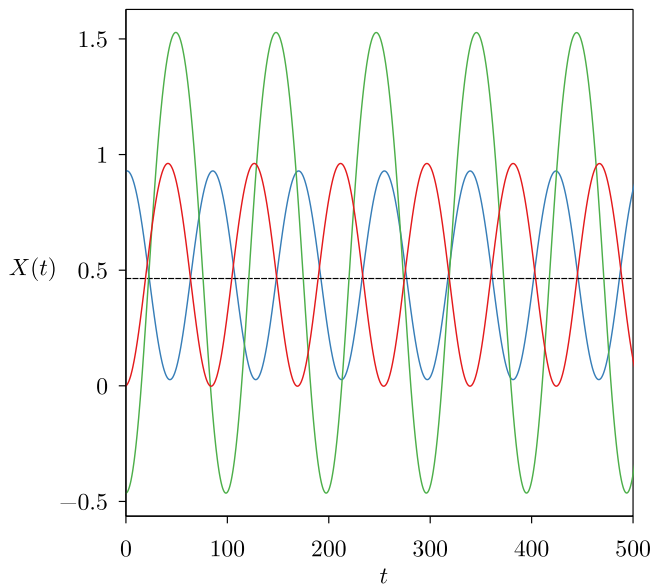}%
	\end{center}
	\caption{Results of the nonlinear dynamics simulation for  $\Phi(x, 0)  = \sech(x-\tilde{x})$ when the initial center of the wavepacket $X(0)=\tx$ is different from the value $x_0$  predicted by the asymptotic analysis. Three curves correspond to $\tx = x_0 \pm x_0$ (blue and red curves, respectively) and to $\tx = -x_0$ (green curve). In all cases $\varepsilon=0.01$. Horizontal dashed line corresponds to $X=x_0$. }
	\label{fig:dyn02}
\end{figure}

Next, we address the situation when the  initial condition   $\Phi(x, 0)  = \sech(x-\tx)$ is situated at a different position  than that prescribed by the asymptotic analysis, i.e., $\tx\ne x_0$. The results plotted in Fig.~\ref{fig:dyn02} show that for small $\varepsilon$ the center of   initially displaced wavepacket performs nearly perfect periodic oscillations around $x_0$, \rev{and the amplitude of the oscillations increases with the increase of the initial displacement $|\tx-x_0|$}. This suggests that even though there is no  authentic   stationary mode existing at $x=x_0$, this asymptotically predicted position still plays an important role in the nonlinear dynamics.  \rev{For extremely large initial positions $\tx$, the initial soliton is situated in an effectively homogeneous medium, because the numerical value of the exponentially decaying potential becomes zero for large $x$. In this situation the periodic oscillations naturally  disappear.}

\begin{figure}
	\begin{center}
		\includegraphics[scale=1.6]{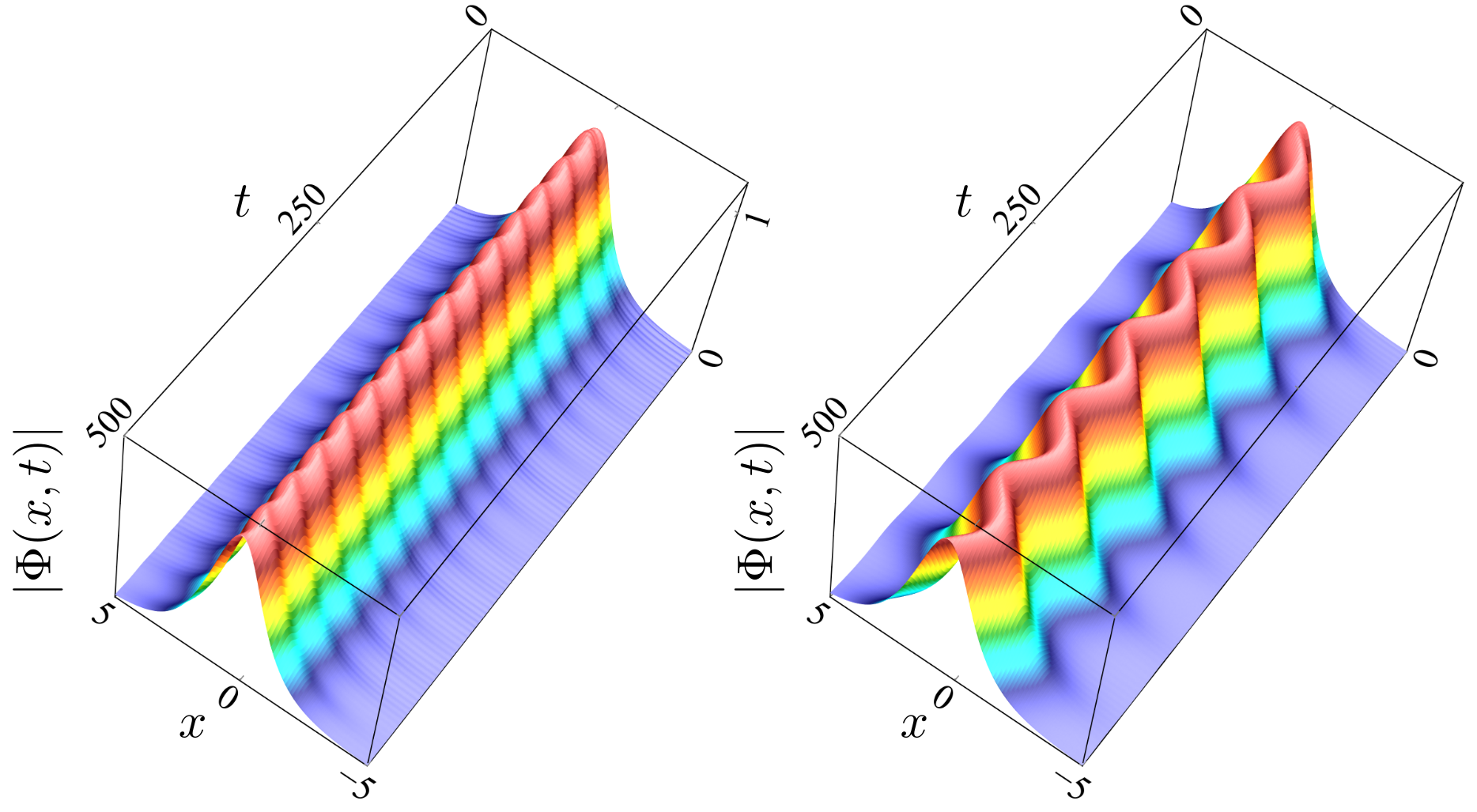}%
	\end{center}
	\caption{Plot of the amplitude $|\Phi(x,t)|$ for  initial condition $\Phi(x, 0)  = \sech(x-\tilde{x})$ when $\varepsilon=0.05$ and  $\tx=x_0$ (left panel) and $\varepsilon=0.01$ and $\tx =-x_0$ (right panel). 	}
	\label{fig:dyn03}
\end{figure}

	Representative plots illustrating   evolution of  the  amplitude $|\Phi(x,t)|$ are presented in Fig.~\ref{fig:dyn03}.

\section{Nonlinear modes for isolated values of $\varepsilon$}\label{sec:SepEps}

\rev{In this section, we   we complement our study by numerical computing the authentic   stationary modes. They are     found  using the extension of the  numerical shooting approach   described above in Sec.~\ref{Sec:matching}. In contrast to the main outcome  of Ref.~\cite{KCKFB19}, we find that stationary  nonlinear modes in a generic W-dW potential of fixed amplitude do  not form a continuous family, but exist only  at isolated values of the  the propagation constant $\mu$. The found authentic nonlinear modes have  complex  two-hump shapes. Their dynamics is unstable.} 

\begin{figure}
	\begin{center}
		\includegraphics[scale=0.9]{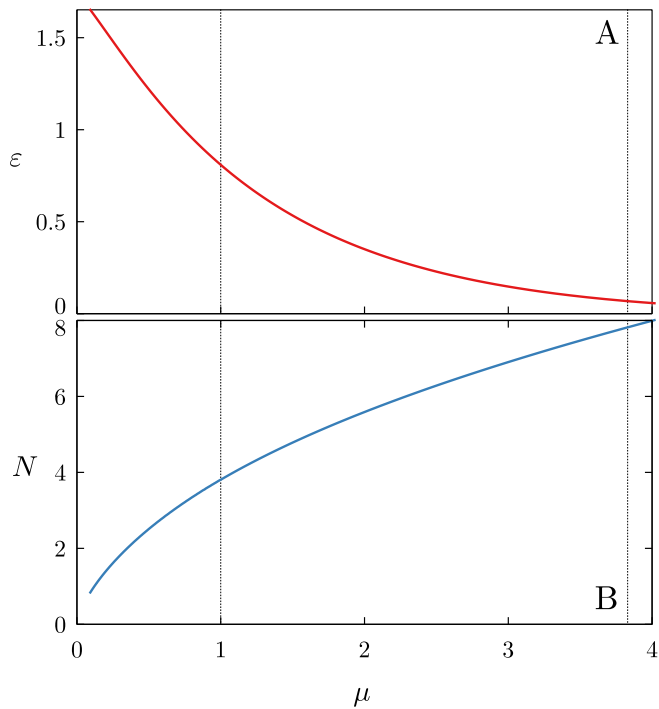}%
	\end{center}
	\caption{Dependencies $\varepsilon(\mu)$ and $N(\mu)$ for authentic   nonlinear modes. Vertical dotted lines correspond to values of $\mu$ for solutions   shown in Fig.~\ref{fig:branch02}. }
	\label{fig:branch01}
\end{figure}

As explained above, for a continuously differentiable stationary  mode $\phi(x)$ the resulting system of matching conditions  (\ref{Eq_modF})-(\ref{Eq_phaseDer})
cannot be generically solved if the values of  $\vep$ and $\mu$ are fixed. However, if $\vep$ or $\mu$ is treated as another unknown, then the system of four equations  is no longer \rev{over}determined, and  a numerical solution can be found by Newton iterations.  A good initial guess for the iterative procedure   is given by the pseudo-mode with $\mu=1$ and $\vep=0.8$ which corresponds to the intersection point P3 (see the corresponding panel in Fig.~\ref{fig:Pseudo}). In Fig.~\ref{fig:branch01} we illustrate a numerically found branch of nonlinear modes in terms of  dependencies $\varepsilon(\mu)$ and $N(\mu)$, where $N=\int_{-\infty}^{\infty} |\phi(x)|^2dx$. We stress that the shown dependencies do not represent a continuous family, because they can exist only  if $\mu$ and $\varepsilon$  are varied simultaneously. For instance, for $\mu=1$, the nonlinear mode can only be found at the isolated value $\varepsilon\approx 0.809$.  

\begin{figure}
	\begin{center}
		\includegraphics[scale=0.9]{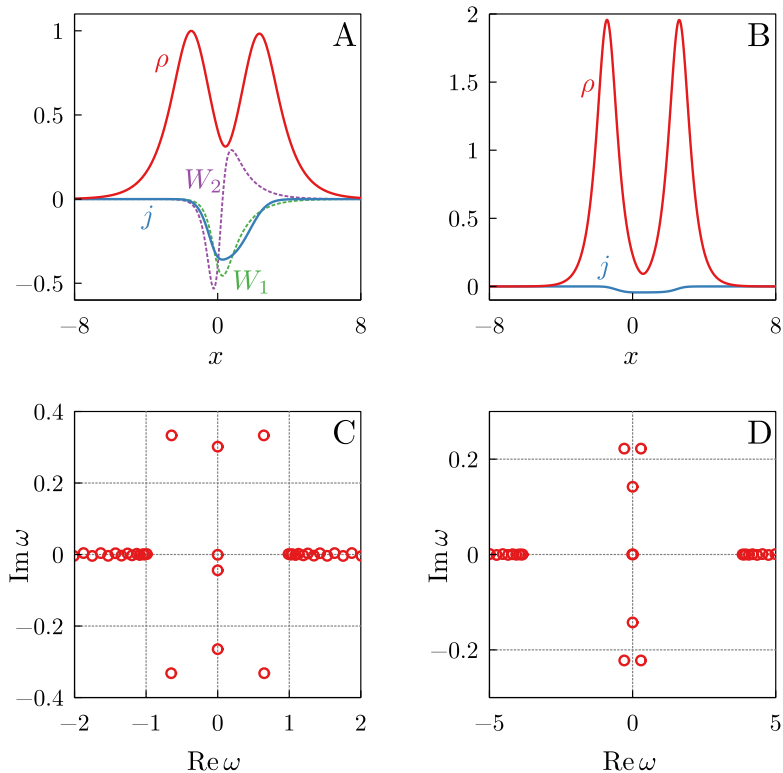}%
	\end{center}
	\caption{
		 Shapes of stationary modes for $\mu=1$, $\varepsilon\approx   0.091$ (A) and $\mu\approx 3.830$, $\varepsilon\approx 0.068$ (B). Red and blue curves correspond to the modulus $\rho(x)  = |\phi(x)|$ and current $j(x)$, respectively. Dashed curves in (A) plot real ($W_1$) and imaginary ($W_2$) parts   of the potential. Lower panels show the linearization eigenvalues.}    	
	\label{fig:branch02}
\end{figure}

Representative shapes of the stationary modes are exemplified in Fig.~\ref{fig:branch02}(A,B) in terms of the amplitude $\rho = |\phi|$ and the current $j(x)$. 
We observe that  the amplitude has a distinctively double-hump shape and features a local minimum approximately at the minumum of the real part of the  potential. 
On the other hand, the current $j(x)$ is dominantly negative, which agrees with the spatial distribution of the gain-and-losses [see the imaginary part of the potential plotted in Fig.~\ref{fig:branch02}(a)].  The maximal negative value of the current is approached at the local minima of the real part of the potential. For small values of $\varepsilon$, the form of the stationary state resembles  a bound state of two    elementary nonlinear modes.  

The eccentric shape of the obtained nonlinear modes suggests that they can hardly be   stable. The instability was indeed confirmed using the linear stability analysis. Following to the standard procedure \cite{Yang2010}, we consider a perturbed solution $\Phi(x,t)  = e^{i\mu t}[\phi(x) + u(x)e^{i\omega t} + \bar{v}(x)e^{-i\bar{\omega}t}]$, where $u(x)$ and $v(x)$ are small perturbations. Linearization of Eq.~(\ref{Eq:Phi}) with respect to $u(x)$ and $v(x)$ gives the linear stability eigenvalue problem
\begin{equation}
\left(\begin{array}{cc}
\partial_{x}^2 - \mu - \varepsilon(W_1 + iW_2) + 4|\phi|^2 & 2\phi^2\\
-2\bar{\phi}^2 & -\partial_{x}^2 + \mu + \varepsilon(W_1 - iW_2) - 4|\phi|^2
\end{array}\right) \left(\begin{array}{c}
u\\
v
\end{array}\right) = \omega \left(\begin{array}{c}
u\\
v
\end{array}\right). 
\end{equation}
Unstable modes correspond to eigenvalues $\omega$ with negative imaginary parts. 

Numerical solution of the linear stability eigenvalue problem   reveals several unstable eigenvalues in the   spectrum, see the eigenvalue portraits in Fig.~\ref{fig:branch02}(C,D). A closer inspection indicates  that, in the contrast with    situation that takes place for  real-valued potentials, $\PT$-symmetric potentials \cite{ZezKon12},  and Wadati potentials \cite{Yang16},   in the case at hand  the  eigenvalues that emerge in   linearization spectra   do not form quartets $(\omega, \bar{\omega}, -\omega, -\bar{\omega})$. This fact provides  another validation of the essentially dissipative nature of W-dW potentials.

\begin{figure}
	\begin{center}
		\includegraphics[scale=1.6]{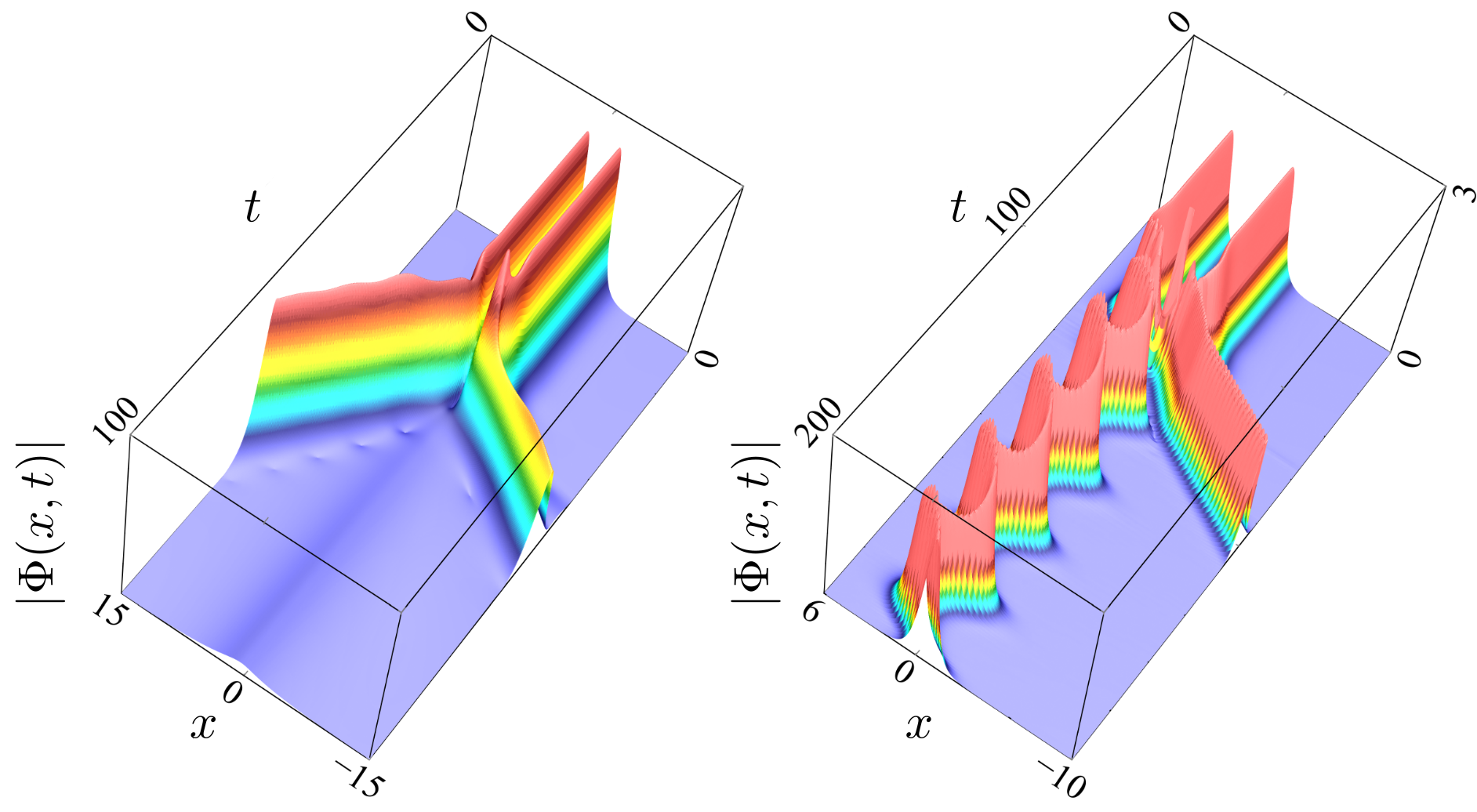}%
	\end{center}
	\caption{Nonlinear dynamics of stationary modes shown in Fig.~\ref{fig:branch02}. 
	}
	\label{fig:branch03}
\end{figure}

Simulating nonlinear dynamics of found stationary modes [exemplified in Fig.~\ref{fig:branch03}], we observe that for larger $\varepsilon$ the mode breaks up into two solitary waves that move in opposite directions and eventually leave the domain where the complex  potential is localized. In the meanwhile, for smaller $\varepsilon$ only one solitary wave escapes, while the second one performs periodic movement which resembles the oscillations of pseudo-modes  observed in Sec.~\ref{sec:Dynamics}.

\section{Conclusion}
\label{sec:concl}
In our study, we have examined    the peculiar features of a  recently discovered class of complex potentials.
More specifically, we have considered the class of {\it W-dW potentials} which by definition have   the form $W(x)=W_{1}(x)+iCW_{1,x}(x)$,  where $W_1(x)$ is a differentiable real-valued function, and   $C$ is a real. It has been   suggested \cite{KCKFB19},
that the nonlinear Schr\"odinger equation (NLSE)   with a W-dW potential  can  support continuous families of stationary solitary-wave nonlinear modes. These objects have been in the focus of the present study.  Assuming that the potential is small, of $\varepsilon$-order, we have employed asymptotic methods to search for the stationary nonlinear modes, seeking them in the form of formal power series with respect to $\varepsilon$. The asymptotic procedure  stops at the terms of the $\varepsilon^2$-order, which  leads us to a conjecture that no  continuous families of nonlinear modes exist in generic W-dW potentials. In order to validate  this hypothesis, we have considered a particular example of the W-dW potential whose real part is a finite-depth   well. The prediction of the asymptotic approach has been confirmed by numerical arguments, because instead of any authentic nonlinear mode we have been able to find only a \textit{pseudo-mode} which solves the equation with $O(\vep^2)$-accuracy.  At the same time, with  numerical simulations of nonlinear dynamics in W-dW potentials, we  have demonstrated that  the  pseudo-modes  can be dynamically robust in small-amplitude W-dW  potentials.  More specifically, the dynamics of pseudo-modes reveals persistent oscillations  of the center-of-mass around the specific position that characterizes the center of the pseudo-mode in the asymptotic expansion.    So, even not being  authentic stationary   nonlinear modes in the mathematical sense, these objects can be regarded as meaningful physical entities. Finally, we have also computed authentic stationary modes which only exist if the parameters of the equation and of the solution itself are tuned precisely. These stationary modes are unstable, and their dynamical instability   reveals  several distinctive   behaviors.

Examples of physically meaningful  ``pseudo-modes'' (or, more generically, ``pseudo-solutions'') are not unheard in the previous literature. For instance, a great number of models where ``asymptotics beyond all orders'' occurs (see \cite{Boyd99} for numerous examples) provide physical objects that cannot be described by idealized mathematical models. One of them is the famous example of $\phi^4$ breather that is nonexistent in mathematical sense \cite{KS87,EKNS88} but that may have ``decay timescale ... longer than the predicted lifetime of the universe''  \cite{Boyd90}. 
\rev{Another   type of pseudo-solutions that is especially well-known in  $\PT$-symmetric potentials  corresponds to  the so-called ghost states, which can be found from the stationary nonlinear equation   with \emph{complex-valued} propagation constants $\mu$ \cite{ghost01,ghost02,ghost03}. Even though  a  ghost state is not  a  true  solution of the time-dependent NLSE, several studies report on  that if the imaginary part of $\mu$ is small, then it can dynamically persist as a metastable  entity, even if the  the coexisting authentic nonlinear modes are unstable. 
	We surmise that the pseudomodes reported herein can be probably also interpreted as ghost states after the consideration is extended to allow   $\mu$ to be complex-valued.} 

To conclude, the search of new   complex potentials that admit continuous families of nonlinear modes      remains a  challenging problem for the future    studies.  Regarding the particular class of  W-dW potentials, we believe  that   an interesting task is related to a more systematic  analysis of   oscillating patterns encountered in the nonlinear dynamics. A more systematic study of  stationary modes in W-dW potentials is also in order. An especially intriguing issue  is the search for    stable stationary modes which can eventually exist in W-dW potentials of the form different from that considered herein. \rev{While only the case of focusing nonlinearity has been considered herein, the behavior of (pseudo)modes can also be addressed   under the the defocusing (repulsive) nonlinearity.}

 \section*{Acknowledgment} 
 
The research by D.A.Z. and G.L.A. is supported by the Russian Science Foundation (Grant No. 20-11-19995).  

 \section*{Data availability statement} 

The data that support the findings of this study are available from the corresponding author upon reasonable request.

\appendix

\section{Calculation of the coefficients for the system (\ref{Eq:LinSys01})-(\ref{Eq:LinSys02}).}\label{Sect:Coeff}

Direct substitution of (\ref{Eq:u_c1})-(\ref{Eq:v_c2}) into  (\ref{Eq:IntCond01})- (\ref{Eq:IntCond02}) yields the system  (\ref{Eq:LinSys01})-(\ref{Eq:LinSys02}) where (for compactness in what follows we write $\int$ instead of $\int_{-\infty}^\infty$, bearing in mind that the integration is always over the whole real axis):
\begin{align}
A_{11}&=\int \left(12 u_0\tilde{u}_1-W_1\right) u_{0,x}^2~dx,\label{A_11}\\[2mm]
A_{12}&=\int  \left(4 u_0\tilde{v}_1 +W_2\right)u_0u_{0,x}~dx,\label{A_12}
\end{align}
\begin{align}
A_{21}&=\int \left(4u_0\tilde{v}_1 -W_2\right)u_0u_{0,x}~dx,\label{A_21}\\[2mm]
A_{22}&=\int \left(4 u_{0}\tilde{u}_1 -W_1\right)u_{0}^2~dx,\label{A_22}
\end{align}
\begin{align*}
F_1&=\int \left(6u_0\tilde{u}_1^2+2u_0\tilde{v}_1^2-W_1\tilde{u}_1+W_2\tilde{v}_1\right)u_{0,x}~dx,\\[2mm]
F_2&=\int \left(4u_0\tilde{u}_1\tilde{v}_1-W_1\tilde{v}_1-W_2\tilde{u}_1\right)u_0~dx.
\end{align*}
Let us simplify these expressions. In the space of rapidly decreasing functions (Schwartz space) define the inner product of $a(x)$ and $b(x)$ as
\begin{gather*}
\langle a(x),b(x)\rangle:=\int  a(x)b(x)~dx.
\end{gather*}
The operators ${\cal L}_2$ and ${\cal L}_6$ are self-adjoint, therefore
\begin{gather*}
\langle {\cal L}_2a(x),b(x)\rangle=\langle a(x), {\cal L}_2b(x)\rangle,\quad \langle {\cal L}_6a(x),b(x)\rangle=\langle a(x), {\cal L}_6b(x)\rangle.
\end{gather*}
Also we make use of the fact that by construction [see equations (\ref{Eq:L6_01})--(\ref{Eq:L2_01}) and (\ref{Eq:u_c1})--(\ref{Eq:v_c2})] 
\begin{gather*}
{\cal L}_6\tilde u_1=W_1u_0,\quad {\cal L}_2\tilde v_1=W_2u_0.
\end{gather*}

1. Consider  $A_{12}$.
\begin{align*}
A_{12}=\int \left(4 u_0\tilde{v}_1 +W_2\right)u_0u_{0,x}~dx=4\int  u_0^2u_{0,x}\tilde{v}_1~dx +\langle u_{0,x},{\cal L}_2\tilde{v}_1\rangle.
\end{align*}
The last term can be transformed   as follows
\begin{align*}
\langle u_{0,x},{\cal L}_2\tilde{v}_1\rangle=\langle {\cal L}_2u_{0,x},\tilde{v}_1\rangle=-4\langle u_0^2u_{0,x},\tilde{v}_1\rangle=-4\int u_0^2u_{0,x}\tilde{v}_1~dx.
\end{align*}
Here we make use of the formula ${\cal L}_6 u_{0,x}=0$ that implies that
\begin{equation}
\label{eq:aux1}
{\cal L}_2 u_{0,x}=-4u_0^2u_{0,x}. 
\end{equation}
Therefore $A_{12}=0$.

2.  Consider $A_{22}$. 
\begin{align*}
A_{22}=&\int (4u_0{\tilde u}_1-W_1)u_0^2~dx=4\int  u_0^3{\tilde u}_1~dx-\langle{\cal L}_6\tilde{u}_1,u_0\rangle\\[2mm]
&=4\int  u_0^3{\tilde u}_1~dx-\langle\tilde{u}_1,{\cal L}_6u_0\rangle.
\end{align*}
Since  ${\cal L}_2 u_{0}=0$ then ${\cal L}_6 u_{0}=4u_0^3$. Therefore
\begin{gather*}
A_{22}=4\int  u_0^3{\tilde u}_1~dx-\langle\tilde{u}_1,4u_0^3\rangle=0.
\end{gather*}

3.  Consider $A_{21}$.
\begin{align*}
A_{21} = A_{12} - 2\int  W_2 u_0 u_{0,x}dx = -   2\la  \cL_2 \tv_1,     u_{0,x}\ra  = 8\int   \tv_1 u_0^2   u_{0,x}dx, 
\end{align*}
where we have again used (\ref{eq:aux1}).

4.  Consider $A_{11}$.
\begin{align}
A_{11} = 12\int  u_0\tu_1u_{0,x}^2 dx  - \int W_1u_{0,x}^2dx  =  12\int   u_0\tu_1u_{0,x}^2 dx  + \int u_0(W_{1,x}u_{0,x} + W_1u_{0,xx})dx=\nonumber\\[2mm]
12\int  u_0\tu_1u_{0,x}^2 dx  + C\int u_{0,x}W_2 u_0 dx + \int  W_1 u_0 (u_0-2u_0^3)dx=\nonumber\\[2mm]%
12\int u_0\tu_1u_{0,x}^2 dx  +  C\la \cL_2\tv_1, u_{0,x}\ra + \la \cL_6 \tu_1, u_0\ra  - 2 \la \cL_6 \tu_1, u_0^3\ra =\nonumber\\[2mm]%
12\int  u_0\tu_1u_{0,x}^2 dx  -4 C\int \tv_1 u_0^2  u_{0,x} dx   +  4\int \tu_1 u_0^3 dx   - 2 \la  \tu_1,  \cL_6u_0^3\ra.
\label{eq:A11}
\end{align}
Straightforward computation yields $\cL_6 u_0^3 = 6u_0 u_{0,x}^2 + 2u_0^3$. This implies that the first,   third and   fourth summands   in (\ref{eq:A11})   annihilate, and we finally obtain
\begin{align*}
A_{11} =   -4 C\int \tv_1 u_0^2  u_{0,x} dx.
\end{align*}

5. Consider $F_2$.
\begin{align*}
F_2 = 4\int  u_0^2\tilde{u}_1\tilde{v}_1 dx -\int W_1 u_0 \tilde{v}_1dx -\int W_2 u_0 \tilde{u}_1 dx =  
4\int u_0^2\tilde{u}_1\tilde{v}_1 dx - \la \cL_6\tu_1, v_1\ra   -\int W_2 u_0 \tilde{u}_1 dx = \\[2mm]
=4\int  u_0^2\tilde{u}_1\tilde{v}_1 dx -\int \tu_1(W_2u_0+4u_0^2\tv_1) dx -\int W_2 u_0 \tilde{u}_1 dx = -2\int W_2 u_0 \tilde{u}_1 dx,
\end{align*}
where we have used the equality $\cL_6 \tv_1 = W_2 u_0 + 4u_0^2 \tv_1$ which can be derived easily.

5. Consider $F_1$. Since its calculation is a bit more involved, we decompose $F_1$ into four summands representing $F_1 = I_1 + I_2 + I_3 + I_4$, where
\begin{align*}
I_1 = 6\int u_0\tu_1^2\uox dx, \quad I_2 = 2\int u_0\tv_1^2 \uox dx,\\[2mm]
I_3 = -\int W_1 \tu_1\uox dx, \quad I_4 = \int W_2\tv_1 \uox dx.
\end{align*}
The calculation proceeds as follows:
\begin{align}
\label{eq:I3}
I_3 = \int u_0( W_{1,x} \tu_1 + W_1 \tu_{1,x}) dx = C\int u_0 W_2 \tu_1 + \la \cL_6 \tu_1, \tu_{1,x}\ra. 
\end{align}
Straightforward differentiation yields $\cL_6  \tu_{1,x} = W_{1,x} u_0  + W_1 \uox  - 12 u_0 \uox \tu_1$, which after substitution in (\ref{eq:I3}) eventually leads to 
\begin{align}
I_3 = 2C\int u_0 W_2 \tu_1 dx - I_3 - 2I_1,
\end{align}
and hence
\begin{equation}
\label{eq:I1+I3}
I_1 + I_3 =  C\int u_0 W_2 \tu_1 dx = -\frac{C}{2} F_2.
\end{equation}
In a similar manner, using that $\cL_2  \tv_{1,x} = W_{2,x}u_0  + W_2 \uox  - 4 u_0 \uox \tv_1$, we deduce
\begin{align*}
I_2+I_4 =  \int \tv_1(2u_0\uox \tv_1 + W_2 \uox)dx  = \frac{1}{2}\int \tv_1(-\cL_2 \tv_{1,x} + W_{2,x}u_0+3W_2\uox) dx=\\[2mm]
\frac{1}{2}\left(-\int W_2 u_0   \tv_{1,x} dx   + \int  \tv_1(W_{2,x}u_0 + 3W_2u_0)dx \right)= \int \tv_1 ( W_{2,x} u_0  + 2W_2 \uox)dx.
\end{align*}
Combining the latter result with (\ref{eq:I1+I3}), we obtain the final expression for $F_1$.

\end{document}